\documentclass{sig-alternate}
\usepackage{algorithm}
\usepackage{algorithmic}
\usepackage{amssymb}

\begin{document}

\title{Color Assessment and Transfer for Web Pages }
\numberofauthors{1}

\author{
%
\alignauthor Ou Wu\\
       \affaddr{National Laboratory of Pattern Recognition, Institute of Automation}\\
       \affaddr{Chinese Academy of Sciences}\\
       \email{wuou@nlpr.ia.ac.cn}
}

\maketitle

\begin{abstract}
Colors play a particularly important role in both designing and accessing Web pages. A well-designed color scheme improves Web pages' visual aesthetic and facilitates user interactions. As far as we know, existing color assessment studies focus on images; studies on color assessment and editing for Web pages are rare. This paper investigates color assessment for Web pages based on existing online color theme-rating data sets and applies this assessment to Web color edit. This study consists of three parts. First, we study the extraction of a Web page's color theme. Second, we construct color assessment models that score the color compatibility of a Web page by leveraging machine learning techniques. Third, we incorporate the learned color assessment model into a new application, namely, color transfer for Web pages. Our study combines techniques from computer graphics, Web mining, computer vision, and machine learning. Experimental results suggest that our constructed color assessment models are effective, and useful in the color transfer for Web pages, which has received little attention in both Web mining and computer graphics communities.
\end{abstract}




\section{Introduction}
Humans generally prefer certain colors over others, and this preference influences a wide range of human behaviors, such as buying cars, choosing clothes, etc. \cite{Palmer:2010}. Colors are also crucial to the success of Web pages and directly affect the perception of their aesthetic and usability \cite{Ling:2002}\cite{Schmidta:2009}. There has been a large amount of work on Web page colors. Kondratova and Goldfarb \cite{Kondratova:2007} conducted studies on color preferences in Web design for a number of countries and identified several country-specific color palettes. Thorlacius \cite{Thorlacius:2007} investigated the effects of visual factors, including colors, typography, and pictures in a Web page, on the users' interaction with that page. Coursaris et al. \cite{Coursaris:2008} studied the effects of color temperature (cool or warm) on Web aesthetics. Their findings suggest that pages with a warm primary color (e.g., red) and a warm secondary color (e.g., orange) are the least aesthetically pleasing. These existing studies focus either on exploring color designing rules or taking colors as an important factor in the evaluation of Web design. So far, little work has been done on the direct evaluation of Web colors, despite the fact that an effective color assessment tool would be useful for many Web-related applications, especially Web appearance design.

Most recently, O'Donovan et al. \cite{Donovan:2011} initiated a pilot work on the construction of assessment models for color compatibility using online data sets that consist of color themes and their associated compatibility ratings. These color themes were created by color experts and the compatibility ratings were the results of votes cast by viewers (users). The constructed regression model proposed by \cite{Donovan:2011} can score a color theme, and their classification model predicts whether or not a color theme is compatible. They proposed several potential image editing applications, such as color theme optimization and color suggestion. The experiments demonstrated the usefulness of their constructed color assessment modes in image editing.

The above work motivated us to investigate the assessment of Web page in terms of color compatibility. Web page designers usually create a small number of colors. Therefore, an intuitive way to apply Web color assessment is to follow the approach proposed by O'Donovan et al. \cite{Donovan:2011}, that is, to obtain the color theme of a Web page, and then to assess its compatibility based on learned models. However, we found this approach inappropriate for direct use with Web pages, as there are obvious differences between a Web page and an image. A Web page, as shown in Fig. 1, has several areas-denoted as temporal part-that display visual contents, such as image, flash, video, etc. The colors found in the temporal part change with the content. The areas outside the temporal part are called fixed. As the colors in temporal parts are changed from time to time and colors in fixed parts are the focus in Web design, this work only assesses the colors of the fixed part. In contrast, studies on image colors do not require distinguishing fixed from temporal parts. Therefore, further study for Web color assessment is required.
\begin{figure*}[!htbp]
\label{f1}
\centering
  \hspace{0in}\includegraphics[width= 12cm, height=1.26in]{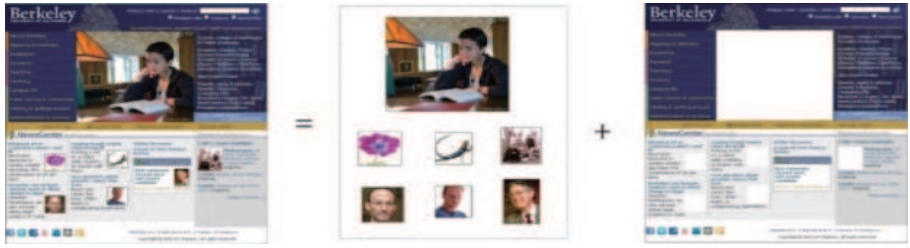}\\
 \vspace{-0.01in}\caption{A Web page (left), its temporal part (middle) and fixed part (right).}
\end{figure*}

The primary goal of our study is to construct a color assessment model for Web pages. Thus, a new construction framework for a Web color assessment model is proposed. Several new algorithms are introduced to address the key steps in the construction framework.

The current study applies the learned color assessment model in a new color editing application, namely, color transfer for Web pages. As in color transfer between images, given a source Web page and a reference Web page, a color transfer algorithm transforms the colors of the source Web page, such that the colors of the transferred source Web page become similar to those of the reference page. An automatical color transfer and assessment framework is presented in this study. This framework can help Web designers to choose Web colors. The new application may increase exchanges between Web mining and computer graphic communities to develop more techniques for computer aided Web design.

\section{Related work}

\subsection{Color assessment}
Current color assessment focuses on assessing the compatibility of the combination of different colors. Color compatibility is related to human color preference, and high color compatibility indicates high color preference. There have been numerous studies on color compatibility assessment. Existing studies can be divided into three categories listed below.

\begin{enumerate}
  \item Single factor-based methods. These studies are based on color compatibility tools, such as color wheels. For example, Goethe \cite{Goethe:1810} pointed out that contrasting colors, i.e., those found on opposite sides of the color wheel, are compatible. Some researchers have compiled color templates based on the color wheel to help color designers.
  \item Multiple factor-based methods. These studies \cite{Lalonde:2007}\cite{Schloss:2011} assess colors depending on hues and other factors, such as saturation, lightness, etc. Unlike the previous category of studies, this kind of work attempts to develop quantitative analysis based on controlled laboratory experiments. However, due to a lack of data, the resulting models sometimes contradict each other.
  \item Learning-based methods. Machine learning is a powerful technique used to construct classification or scoring models based on a larger amount of training data. As discussed earlier, O'Donovan et al. \cite{Donovan:2011} conducted a pilot study to learn a classification/scoring model that quantitatively rates the compatibility of color themes using online color theme-rating data.
\end{enumerate}

Some other studies utilize the abovementioned color assessment theories to guide color selection and enhancement. For example, Cohen-or et al. \cite{Cohen-Or:2006} proposed a color harmonization method based on color wheel. Lalonde et al. \cite{Lalonde:2007} identified realistic images using color compatibility assessment results. There are online color assessment systems for Web pages such as \cite{Checkmycolors}; however, these systems merely check local regions and focus on the color contrast. A system that assesses the overall colors does not currently exist.

\begin{figure*}[htbp]
\label{f1}
\centering
  \hspace{0in}\includegraphics[width= 16.6cm, height=1.86in]{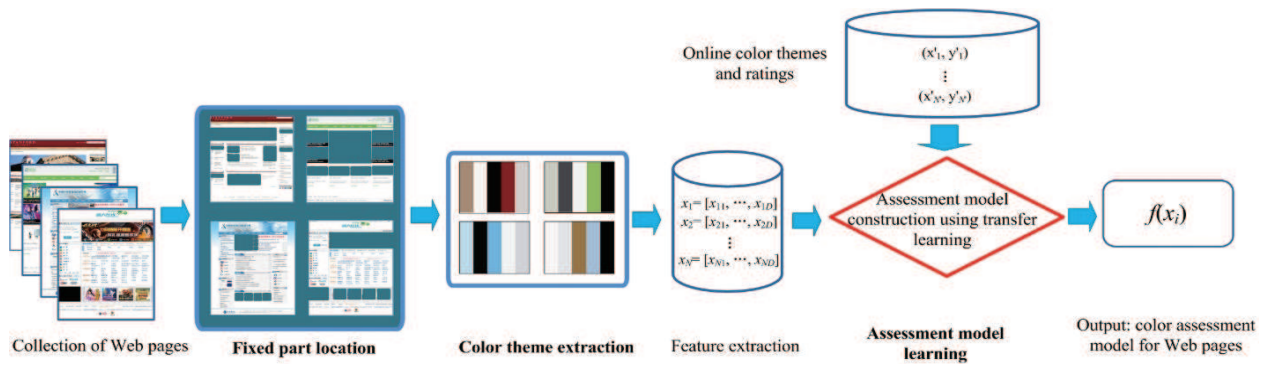}\\
 \vspace{-0.01in}\caption{The steps for the construction of a color assessment model for Web pages.}
\end{figure*}
\subsection{Color transfer}
Color transfer is an image editing technique that keeps the scene from a source image and applies the color style of a reference image \cite{Dong:2010}. Reinhard et al. \cite{Reinhard:2001} proposed the first color transfer algorithm based on the mean and standard deviation of the color values in the source and reference images. The method is efficient but in most cases, artifacts are created when the source and reference images have different color distributions. In response, researchers have developed numerous solutions that transfer the colors of pixels locally \cite{Pitie:2007}\cite{Tai:2005}. Color transfer has been successfully applied in photo appearance enhancement and movie post-processing. Little headway has been made on color enhancement for Web pages, despite the fact that color is also one of the main concerns of Web design.

\subsection{Web appearance mining}

 A large number of recent studies \cite{A-Cai-TR}\cite{Wu:2011} have given attention to the analysis of visual appearance of Web pages. Cai et al. \cite{A-Cai-TR} introduced a visual-based page segment algorithm (VIPS) to extract the visual structure of a Web page. Wu et al.\cite{Wu:2011} proposed an automatic approach for determining whether or not a page is aesthetic. These results are summarized into a new Web mining division, i.e., Web appearance mining, as the (partial) analysis target is the appearances of Web pages. Web appearance mining focuses on discovering useful information (e.g., useful content blocks in \cite{A-Cai-TR}) based on Web appearance. The current study also focuses on the appearance of a Web page. Therefore, part of the present work belongs to Web appearance mining.

\section{Overview of the framework}

Our proposed approach falls under a conventional machine learning approach, involving extracting features and then learning the assessment model (a regression function). The proposed framework is summarized in Fig. 2. There are three main challenges facing the model construction as discussed below.

\begin{itemize}
  \item Fixed part location. The location of the fixed part should be intuitively determined through source code analysis. However, integrating the processing of source code may be impractical for the assessment model because the source codes of many Web pages are irregular and incomplete.
  \item Color theme extraction. Color theme extraction aims to obtain the major colors of a Web page; therefore, minor colors should be excluded.
  \item Assessment model learning. The training data should consist of color themes of Web pages and their respective ratings. Unfortunately, little such training data exists. The online color theme-rating data used by \cite{Donovan:2011} can be utilized as the training data. However, according to the machine learning theory (analyzed in detail in Section 6), the online data set is inappropriate for use because online color themes and Web color themes have different distributions.
\end{itemize}

To address the above challenges, a series of new algorithms are proposed. Processing source codes to locate the fixed part of a page can be avoided using a computer vision method. In color theme extraction, a new clustering algorithm is introduced to discover major colors as well as to exclude color outliers. Transfer learning is introduced into model learning to reweight the online training data and adapt the data distribution to that of Web color themes. Likewise, we adopt a machine learning strategy called ensemble learning to improve the generalization capability of our color assessment model.

Our proposed framework differs from the approach used in \cite{Donovan:2011} in two aspects: (1) in our framework, a Web page should be preprocessed to locate the fixed part, and (2) a transfer learning strategy is used to utilize the plentiful labeled resources of color themes available online. Sections 4, 5 and 6 explain the technical details of the three core steps, namely, fixed part location, color theme extraction and model learning, respectively.
\begin{figure*}[htbp]
\label{f1}
\centering
  \hspace{0in}\includegraphics[width= 14.6cm, height=1.66in]{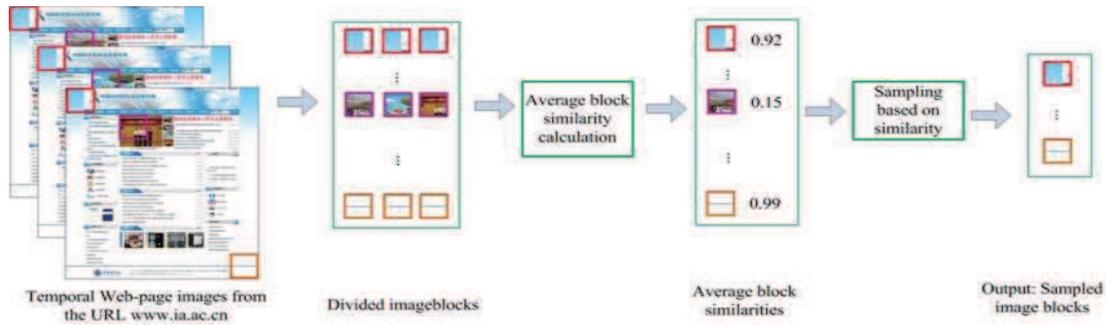}\\
 \vspace{-0.01in}\caption{The main steps of block sampling-based fixed part location.}
\end{figure*}
\section{Fixed part location}
Before a Web page is processed, it is transformed into an image, which is referred to as the Web-page image. When there is no ambiguity, it is still called Web page for the sake of brevity. The process of locating the fixed parts is a Web structural mining problem. \emph{However, many existing Web structural mining algorithms rely heavily on the source codes, limiting their application, given that the source codes of many Web pages are non-standard and noisy}. The visual-based page segment algorithm \cite{A-Cai-TR}, one of the most famous Web page segmentation methods, is likewise difficult to use when the source codes of Web pages are too complex or non-standard. Hence, this study does not utilize these structural analysis algorithms. Instead, we leverage computer vision techniques to locate fixed parts, directly running on the transformed Web-page images. Three new method are proposed in this work, namely, block sampling-based, salience map-based, and image synthesize-based methods. These new methods are independent of source codes and are, therefore, more available in real applications. Experiments show that the first method is more effective and efficient, so in our final work only this method is used and the latter two methods are introduced in the appendix part.

\subsection{Block sampling-based method}
Our method generates a set of image blocks from a sequence of temporal Web-page images giving a URL\footnote{Temporal Web pages of a given URL can be obtained from http://web.archive.org/ automatically.} as shown in Fig. 3. There are three steps involved. (1) Each of the temporal Web-page images is divided into $N_1*N_2$ blocks. Assuming that there are I temporal Web-page images, then for each block position we obtain I image blocks after division. (2) We calculate the similarity between a block of the first Web-page image and the corresponding blocks of the successive temporal Web-page images. As a result, I-1 similarities are obtained for each block. These similarities are then averaged and normalized into $[0, 1]$. The similarity between two image block is calculated based on the earth mover distance (EMD) \cite{Rubner:2000} between the color histograms. Assume that the EMD is $d$, then the similarity is defined as $exp(-d)$. (3) The blocks are sampled according to their average similarities and then used to construct the set of image blocks. The sampling strategy ensures that colors in fixed parts are sampled with higher probabilities than those in temporal parts.

The results of this method on an exemplar temporal Web-page image are shown in Fig. 4. The left image in Fig. 4 shows the fixed part (not including the black areas) and the right one shows the sampled blocks (not including the black areas). Both $N_1$ and $N_2$ are set to 40. The blocks in the fixed part are more sampled than those in the temporal parts. The sampled blocks do not affect the major colors of the fixed part, although some blocks in the fixed part are also not sampled as well. Most temporal parts are not sampled (black areas).

\begin{figure}[tbp]
\label{f1}
\centering
  \hspace{0in}\includegraphics[width= 8.36cm, height=1.66in]{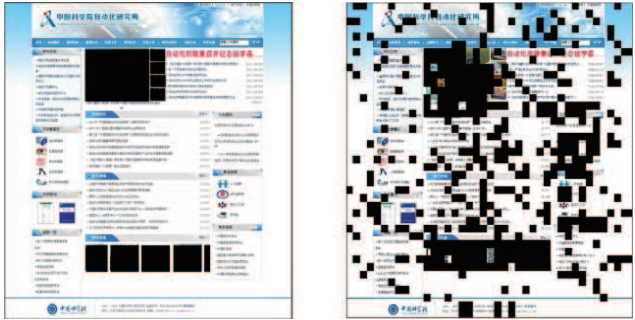}\\
 \vspace{-0.01in}\caption{A Web page's fixed part (left) by manually determined, and the sampled blocks of the Web page (right) by our method. The black boxes are not included.}
\end{figure}

\section{Color theme extraction and features}

Our goal for Web page color theme extraction is somewhat different from that of color theme extraction in O'Donovan et al. \cite{Donovan:2011}, whose goal is to extract a color theme that can represent the colors while also being highly rated. This means that the color theme extracted by \cite{Donovan:2011} may not be the most representative color theme for an image. In contrast, our goal is restricted to the extraction of a color theme that can represent the colors of a Web page as much as possible, whether or not its ratings are high. As such, color theme extraction becomes a pure data clustering problem. A Web page usually contains a large number of different colors. Some colors (e.g., range red colors in the page in Fig. 4 (left)) occupy a very small proportion. These colors have almost no effect on viewers' visual perception of the colors of whole pages. Therefore, they should be considered as color outliers. Nevertheless, conventional clustering techniques such as K-means are sensitive to outliers. To deal with this problem, we introduce the outlier-aware clustering algorithm\cite{Forero:2011} into this work.

\subsection{Outlier-aware clustering}
Data clustering was used in \cite{Donovan:2011} to extract color themes for images. Forero et al.\cite{Forero:2011} proposed the following clustering model which explicitly accounts for outliers:

\begin{figure}[t]
\label{f1}
\centering
  \hspace{0in}\includegraphics[width= 8.56cm, height=1.56in]{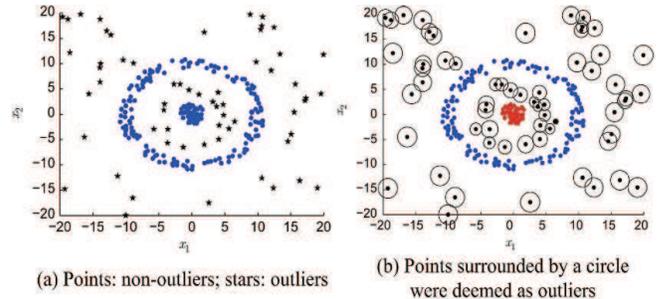}\\
 \vspace{-0.01in}\caption{Results of the outlier-aware clustering algorithm on a synthetic data set (\cite{Forero:2011}).}
\end{figure}

\begin{equation}
\mathop {{\rm{min}}}\limits_{M,O,U} \sum\limits_{i = 1}^n {\sum\limits_{k = 1}^K {u_{ik} } ||x_i  - m_k  - o_i ||_2^2  + } \lambda {\rm{ }}\sum\limits_{i = 1}^n {||o_i ||_2 }
\end{equation}
where $x_i$ is a data point; $u_{ik} = 1$ if $x_i$ belongs to $k$-th cluster and $u_{ik} = 0$ otherwise; $m_k$ represents a centroid; $\delta()$ denotes the indicator function; $\lambda \geq 0$ is an outlier-controlling parameter, such that the higher the value of $\lambda$, the less the number of the points detected as outliers. When $\lambda \rightarrow \infty$, all the data are deemed outlier-free and the outlier-aware clustering equals to K-means. 
The detailed steps to solve Eq. (1) can be referred to as the Robust K-means algorithm proposed by \cite{Forero:2011}. In this work, $K$ equals 5 and $\lambda$ is set to 70.

Figure 5 shows the results of the above outlier-aware clustering algorithm on a synthetic data set (taken \cite{Forero:2011} from directly)\footnote{In the supplementary material (http://minus.com/lwD8XXA4mAsEY), we compare the above algorithm with K-means in color theme extraction for images. Results suggest that the above algorithm can extract more representative colors from the original images. Please kindly check it.}. The two clusters are correctly obtained while all the outlier data are correctly detected by the algorithm.

\subsection{Color theme extraction with fixed part location}
In this stage, five representative colors are obtained by applying outlier-aware clustering to the colors of the fixed part of a Web page obtained by the proposed block sampling-based method. Pixels in some blocks may be sampled more than once, a weighted outlier-aware clustering algorithm is required. Equation. (1) can then be transformed:
\begin{equation}
\mathop {{\rm{min}}}\limits_{M,O,U} \sum\limits_{i = 1}^n {\sum\limits_{k = 1}^5 {w_i u_{ik} } ||x_i  - m_k  - o_i ||_2^2  + } \lambda {\rm{ }}\sum\limits_{i = 1}^n {||o_i ||_2 }
\end{equation}
where $w_i$ is the sampled times of the $i$-th pixel. The solution of Eq. (2) is similar to that of Eq. (1) with a trivial modification.

The five representative colors obtained should be combined from left to right to form a color theme. There are $5!=120$ possible left to right combinations of the five colors. Given that the differences between two adjacent colors in a color theme affect the rating of the color theme \cite{Donovan:2011}, it is inappropriate to randomly set the relative spatial positions of the five representative colors. A simple method is proposed to capture partial spatial information of the five representative color. Once color clustering is completed, we calculated the pair-wise distances between clusters. The pair-wise distance between two clusters is the average of the pair-wise distances of the positions of the pixels in the two clusters, thereby indicating the spatial information of the five colors. Finally, we selected an optimal color combination from the 120 possible combinations, such that the relative positions agree with the pair-wise distances as much as possible.

Figure 6 shows three exemplar results of the extracted color themes based on different methods: K-means (in (a)), outlier-aware method with salience map-based location (in (b)), outlier-aware method with image synthesize-based methods\footnote{Please refer to Appendix for more details.}, and outlier-aware method with block sampling-based location (in (d)). The color themes from outlier-aware clustering + block sampling-based location are more insensitive to the colors of the temporal part than other three methods. For instance, in Fig. 6(2), only the color theme in (d) does not contain the purple color which does not appear in the fixed part.

\begin{figure}[t]
\label{f11}
\centering
  \hspace{0in}\includegraphics[width= 8cm, height=5.08in]{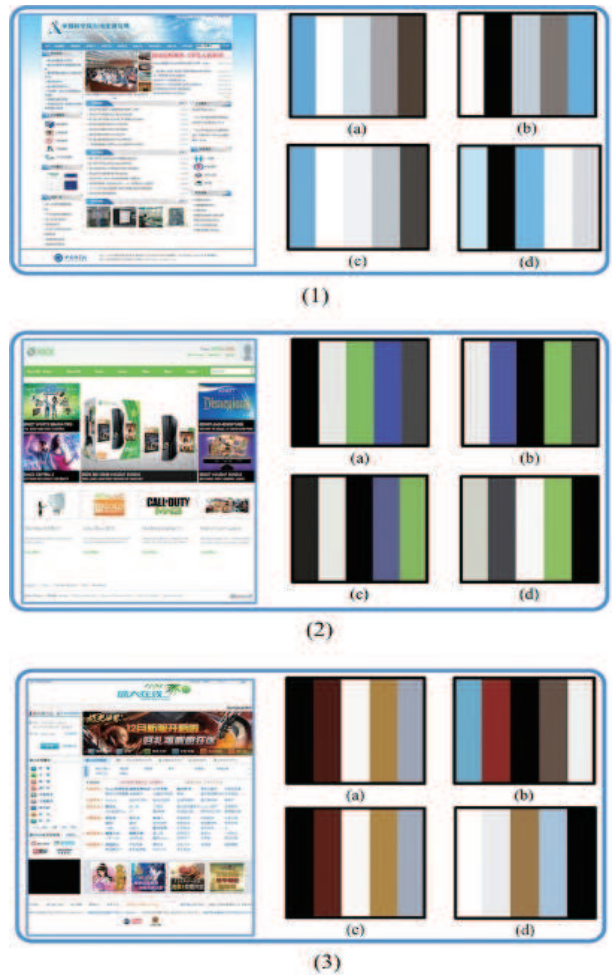}\\
 \vspace{-0.01in}\caption{Color theme extraction for Web pages: (a) K-means, (b) Outlier-aware clustering + Salience map-based fixed part location, (c) Outlier-aware clustering + Image synthesize-based fixed part location, and (d) Outlier-aware clustering + Block sampling-based fixed part location.}
\end{figure}

\subsection{Features}
 The features proposed in \cite{Donovan:2011} are leveraged in this work with a slight modification. The mean values are weighted by the proportions of pixels in each of the five colors in the color theme obtained by the clustering algorithm. Finally, a feature vector with 334 dimensions is constructed to represent the color theme of a Web page. This vector contains comprehensive factors related to color compatibility.

\section{Assessment model learning}

\subsection{Problem description}
Let $P_W$ denote the distribution of color themes of Web pages. Intuitively, $P_W$ mainly depends on the distribution of Web pages. Let $P_O$ denote the distribution of online color themes created by color experts. Intuitively, $P_O$ mainly depends on color experts who created them. Then, we have
$P_W\neq P_O$. Based on a machine learning perspective, if the distribution of future (test) data, i.e., $P_W$,  is different from training data, i.e., $P_O$, the learned model usually has poor generalization capability.

The learning problem is then described as follows. \emph{Given a set of features ($X = {x_1, \cdots, x_N}$) of color themes extracted from a collection of Web pages, the underlying distribution of $X$ is $P_W$. There exists a set of features $X^{'} = \{x^{'}_1, \cdots, x^{'}_{N^{'}}\}$ of online color themes and the associated user ratings $Y^{'} = \{y^{'}_1, \cdots, y^{'}_{N^{'}}\}$. The underlying distribution of $X'$ is $P_O$. $P_W$ is different from $P_O$. Then, how to learn a color assessment model for Web pages using $X$, $X^{'}$, and $Y^{'}$.}

\subsection{Learning with transfer learning}
If the online and Web color themes are denoted as the source and target domains, respectively, the above problem is that learning an assessment model in the target domain with the help of the source domain which has a large amount of labeled training data. This is just a standard transfer learning problem. Huang et al. \cite{Huang:2006} proposed an effective transfer learning algorithm that reweights the data in the source domain, in order to adapt the distribution of the re-weighted data to approach to the distribution of the target domain. Let $\kappa$ be a radial basis kernel function between two samples.
\begin{equation}
\begin{array}{l}
\Phi_{ij}  = \kappa(x'_i ,x'_j ) \quad (x'_i ,x'_j \in X') \\
\phi _i  = \frac{N'}{{N}}\sum\limits_{j = 1}^{N} {\kappa(x'_i ,x_j )} \quad (x'_i \in X', x_j \in X)\\
 \end{array}
 \end{equation}
$\Phi$ can be viewed as a measure of the similarities among source data, and $\phi$ measures the similarities among source and target data. Let $\beta = \{\beta_i\}(i=1,\cdots,N')$ which denotes the weights of the samples. $\beta$ can be obtained by solving the quadratic problem below.
\begin{equation}
\begin{array}{l}
 {\rm{min }}\frac{1}{2}\beta ^{\rm{T}} \Phi\beta  - \phi ^{\rm{T}} \beta  \\
 s.t.{\quad\rm{ }}\beta _i  \in {\rm{[0, }}B{\rm{]  \quad and  \quad}}\left| {\sum\limits_{i = 1}^{N'} {\beta _i  - N'} } \right| \le N'\varepsilon  \\
 \end{array}
\end{equation}
$B$ is set to 1000 and $\varepsilon$ is set to 1 in this work. Equation (4) shows that a larger similarity between a source data point and all the target data points results in a larger weight for that source data point. This is reasonable because domain adaptation aims to reweight the distribution of source data into target data.

After determining $\beta$, a color assessment function can then be obtained using a regression algorithm. This study uses the LASSO regression method because of its good performance reported in \cite{Donovan:2011}. Combined with the weights, LASSO can be written as below.
\begin{equation}
\mathop {{\rm{min}}}\limits_{a,b} \sum\limits_{i = 1}^N {\beta _i (ax_i  + b - y_i )^2 }  + \lambda \left\| a \right\|_1
\end{equation}
where $a$ is the weight vector for features and $b$ is a constant. The above equation can be solved by a convex optimization algorithm \cite{Friedman:2009}.

\begin{algorithm}[h]
    \caption{Ensemble-based transfer learning}
    \label{A-2}
    \begin{algorithmic}[1]
\REQUIRE ~~ $X$, $X^{'}$, $Y^{'}$, $\lambda$, $B$, $\varepsilon$, $L$;
\ENSURE ~~ $a$, $b$;
\STATE Calculate $\beta$ based on Eq. (4).
\FOR{(int $l=1$; $l\leq L$; $l++$)}
    \STATE Generate a new null training set $T_l$.
\FOR{(int $i=1$; $i\leq N^{'}$; $i++$)}
\STATE Generate a random number $d$ in $[0, Max(\beta)]$.
\IF{$d < \beta(i)$}
\STATE Insert $\{x^{'}_i, y^{'}_i\}$ into $T_l$.
\ENDIF
\ENDFOR
\STATE Generate $a_l$ and $b_l$ by optimizing Eq. (5) on $T_l$.
\ENDFOR
\STATE Return $a=sum(a_l)/L$ and $b=sum(b_l)/L$.
\end{algorithmic}
\end{algorithm}

\subsection{Improvement by ensemble learning}
Ensemble learning is another machine learning strategy that can improve the generalization capability of a model. Therefore, this section utilizes ensemble learning to further improve the generalization capability. Specifically, a new ensemble strategy is proposed by combining the weights obtained from transfer learning and bagging strategy \cite{Dietterich:2000}. A new training set is produced in each run of bagging by sampling the data in the source domain according to their weights obtained from Eq. (4). The generated new training set is then used to learn a regression function by LASSO shown in Eq. (5). Finally, all the learned functions are summed as the final scoring model for color assessment. Steps for the ensemble-based transfer learning are shown in Algorithm 1.
\begin{figure}
\label{f1}
\centering
  \hspace{0in}\includegraphics[width= 8.66cm, height=1.32in]{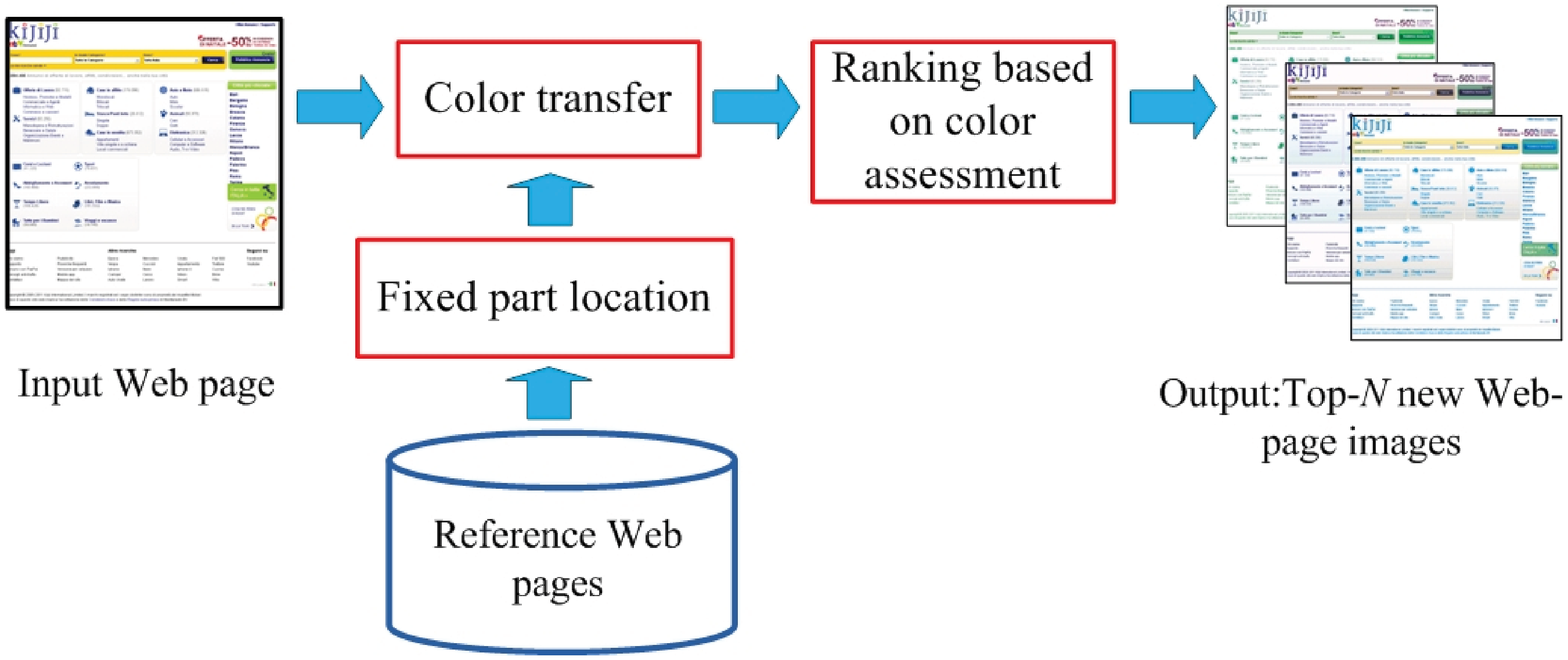}\\
 \vspace{-0.01in}\caption{The main steps of automatic color transfer for Web pages.}
\end{figure}
\begin{figure*}[htbp]
\label{f1}
\centering
  \hspace{0in}\includegraphics[width= 16.16cm, height=1.36in]{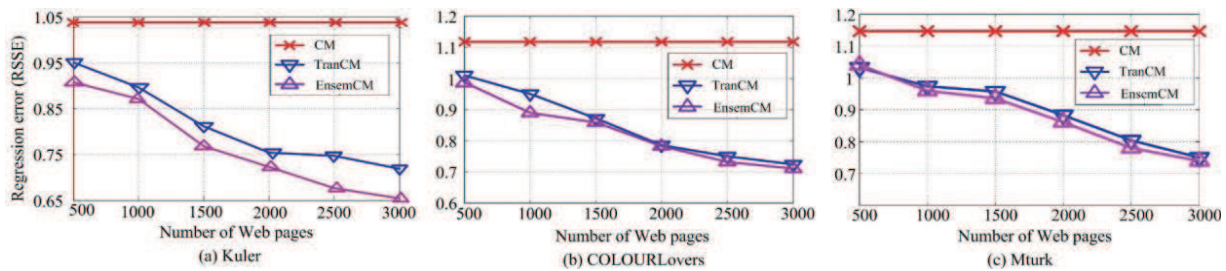}\\
 \vspace{-0.01in}\caption{The regression errors of the three competing assessment models for Web color assessment.}
\end{figure*}

\begin{figure*}[htbp]
\label{f1}
\centering
  \hspace{0in}\includegraphics[width= 15.88cm, height=1.216in]{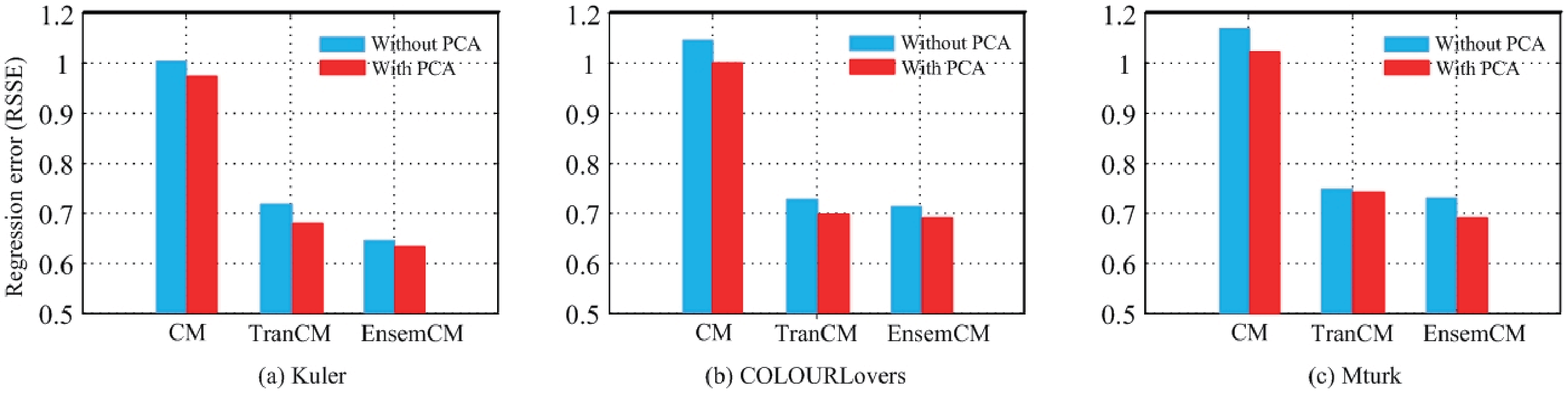}\\
 \vspace{-0.01in}\caption{The regression errors of the three competing assessment models without or with PCA.}
\end{figure*}
\section{Application to color transfer}
The learned color assessment model in Section 6 is applied into color transfer for Web pages, a relatively new application. Given a source Web page and a reference page, the first step is to locate the fixed parts of both source and reference pages. In our work, the fixed part of the reference page, whose colors we aim to transfer into, is obtained using the methods proposed in Section 4. For the source page, users manually identify the fixed part.

Our color assessment and transfer for Web pages is summarized as an approach presented in Fig. 7. The approach automatically transfers the colors of an input Web page into the reference pages in the collections. The top-$N$ transferred Web pages with higher color scores are then selected. Our proposed Web color transfer approach differs from conventional image color transfer in two aspects: (1) our approach includes a color assessment step, which can help users obtain high-quality transfer results, and (2) only the colors in the fixed parts of reference Web pages are considered, so the proposed fixed part location method should be used. On the contrary, in image color transfer, there is no a corresponding step.

The three key components in the proposed application are detailed below.
\begin{itemize}
  \item Constructing a reference Web page collection. This step collects representative Web pages for color transfer use.
  \item Transferring colors based on the Web page database. This step takes turns to select one page in the Web page collection as the target Web page. The fixed part of the target page is extracted and color transfer is then performed.
  \item Ranking the transferred pages based on color assessment. This step orders the transferred Web pages. A preferred Web page list can be obtained according to the color scores of the transferred pages. Finally, Top-$N$ pages are output where $N$ can be set by the users.
\end{itemize}

The above components are explained in the four points:
\begin{enumerate}
  \item In color transfer, it is unnecessary to take all the pages in the reference collection as reference pages. A more feasible solution is to only take similar Web pages as references. The similarity among pages can be defined according to application context. For example, if two Web pages are both the homepages of enterprisers, they are considered similar. The similarity can also be measured by their structures. Web content and structural mining techniques may help in the measurement.
  \item Color transfer can be performed using any of the existing algorithms such as those of \cite{Reinhard:2001} and \cite{Pitie:2007}. The current study does not intend to explore the optimal algorithm among the existing color transfer algorithms. In our work, we directly leverage the work proposed by Piti{\'e} et al. [2007] as the basic color transfer algorithm.
  \item \textbf{Ideally, in the step of color assessment rank, a visual quality evaluation algorithm should be used to assess the transferred pages. } However, studies on visual quality for Web pages are still in infancy. Compared to the large number of online color rating data, the available rating data for the visual quality of Web pages is very limited \cite{Wu:2011}. Hence, in this study, we rate the color transfer results still based on our color assessment model which is learned from online color rating data.
  \item The elements of the Web page collection can be any visual objects, such as images, paint, etc. The kind of visual presents used in the collection depends on the application context or user choices.
\end{enumerate}


\section{Experiments}
This section aims (1) to evaluate the proposed framework for the construction of an assessment model for Web page colors, and (2) to investigate whether the proposed color transfer approach is useful or not. Therefore, the proposed color theme extraction and color assessment model construction methods are evaluated in Sections 8.1 and 8.2, respectively; several examples of color transfer for Web pages with an online user study are present in Section 8.3.

In color theme extraction and color assessment, we chose homepages as our test data and collected \textbf{500} homepages, mainly from companies, universities, governments, and personal sites. A total of nine graduate students from the laboratory, specifically six males and three females, are invited to label the collected pages. Each participant is allowed to view one page within 5 seconds and assess the color design of the page from the five rating scores (1, 2, 3, 4, and 5). Here, ``1" means very bad, and ``5" means very good. After human labeling, each page obtained nine scores. The average of these scores is taken as the color score.
\begin{figure*}[htbp]
\label{f1}
\centering
  \hspace{0in}\includegraphics[width= 16.8cm, height=1.86in]{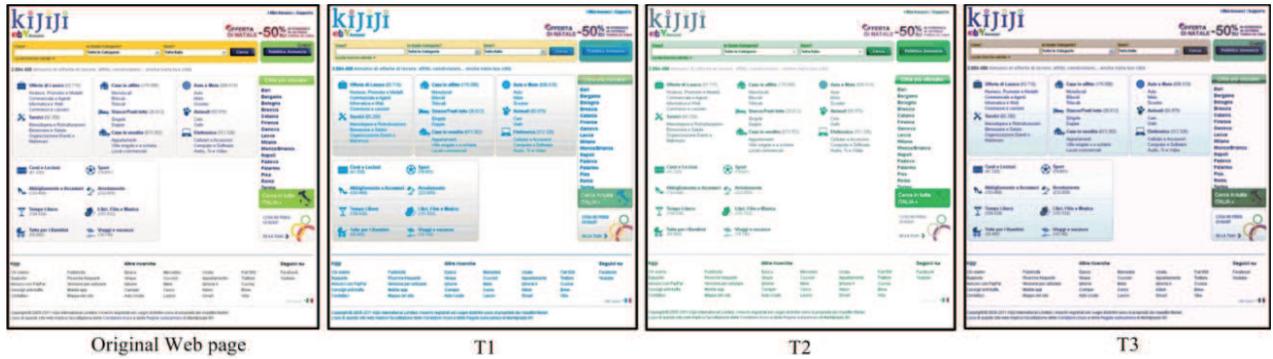}\\
 \vspace{-0.01in}\caption{The original Web page and three transferred Web pages. In our online user study, the average rating scores by 110 users of the two transferred pages (T2 and T3) are higher than that of the original page.}\vspace{0cm}
\end{figure*}
\subsection{Results on color theme extraction}
The four color theme extraction methods (K-means, outlier-aware clustering + salience map-based location, outlier-aware clustering + image synthesized-based location, and outlier-aware clustering + block sampling-based location) shown in Fig. 6 are compared on randomly selected 100 collected test Web pages, whose fixed parts are manually determined. Let $x_i$ be a pixel belonging to the fixed part of a Web page, and $M = \{m_1, m_2, m_3, m_4, m_5\}$ be the color theme of that page obtained by a extraction method. $M$ can be evaluated by the average within-cluster sum (aCS) of squares below.
\begin{equation}
aCS = \frac{1}{K}\sum\limits_{k = 1}^K {\sum\limits_{x_i \in m_j (m_j \in M) } {\left\| {x_i  - m_j } \right\|_2} }
\end{equation}
where $x_i \in m_j$ means that $m_j$ is the closest color to the $i$-th pixel among the five colors in the color theme. A lower aCS value indicates a better color theme ($M$).

On each page, each method is repeated 5 times as the underlying clustering algorithm is sensitive to initial points. The average of aCS on the 100 pages for the four competing methods are 6.357, 5.2663, 4.943, and 4.245 in the CIELab space, respectively. Our introduced outlier-aware method with block sampling-based fixed part location achieves the minimum mean of aCS values, indicating its best performance among the competing methods. The following experiments will merely use the sample-based method in the fixed part location.

\subsection{Results on Web color assessment}
The assessment models below are compared.
\begin{itemize}
  \item The color assessment model proposed by \cite{Donovan:2011}, which is denoted as CM in this paper, and is directly learned from the online color theme-rating data;
  \item The color assessment model constructed by the proposed framework (shown in Fig. 2), which is based on the transfer learning shown in Eq. (5), and is denoted as TranCM.
  \item The color assessment model constructed by the proposed framework (shown in Fig .2), which is based on the ensemble-based transfer learning shown in Algorithm 1, and is denoted as EnsemCM.
\end{itemize}

A source data set and a target data set should be prepared for the learning step in the construction of the TranCM and EnsemCM models. All the three online data sets (Kuler, COLOURLovers, and Mturk) used in \cite{Donovan:2011} are used to create the source data. We randomly selected 3000 samples from each online data set to create a new data set as the source data. For the target data set, we collected $500*n$ ($n$=1, 2, 3, 4, 5, and 6) Web pages based on the Alexa rankings \cite{Alexa} by deleting pages that are duplicated with nearly the same URLs (e.g., www.google.com and www.google.gr). After compiling the source data set and target data set and then extracting their color-theme features, the TranCM and EnsemCM models can be obtained using the corresponding learning algorithms.

Each of the three competing color assessment model is run on the features of the 500 collected test Web pages to score these test pages. The corresponding values of residual sum-of-square error (RSSE) are recorded. The above process is then repeated five times, after which the average RSSE values are recorded. The parameters are searched by cross validation. For EnsemCM, $L$ is set to 50.

Figure 8 shows the results of the three competing models on the 500 Web pages in terms of residual square sum of error (RSSE).  Each image in Fig. 8 represents the results when the source data are created by the Kuler, COLOURLovers, and Mturk, respectively. A total of 500*$n$ ($n$=1, 2, 3, 4, 5, and 6) Web pages are used as the target data. When the number of target data is increased, the RSSE values (regression errors) are decreased. This is reasonable because the increasing number of target data causes the distribution of target data to more affect the weights of the source data. The proposed transfer learning based model (TranCM) outperforms the existing model (CM), while the proposed ensemble transfer learning based model (EnsemCM) achieves the best results over the three data sets. These results suggest that transferring the online color data to assess the colors of Web pages is helpful. The introduced ensemble learning is useful in the transfer.

As some features are correlated, thus, we further applied principal component analysis (PCA) \cite{Cao:2003}, which is a feature reduction technique, on the involved data sets in order to de-correlate the features. Figure 9 shows the results of the performances of the competing models when PCA is used. A total of 3000 samples are randomly selected for each online data set. All 3000 samples are used for Web data. The RSSE values of all the competing models are reduced, indicating that de-correlation can improve the assessment performance.

\begin{figure*}[htbp]
\label{f2}
\centering
  \hspace{0in}\includegraphics[width= 17.66cm, height=1.386in]{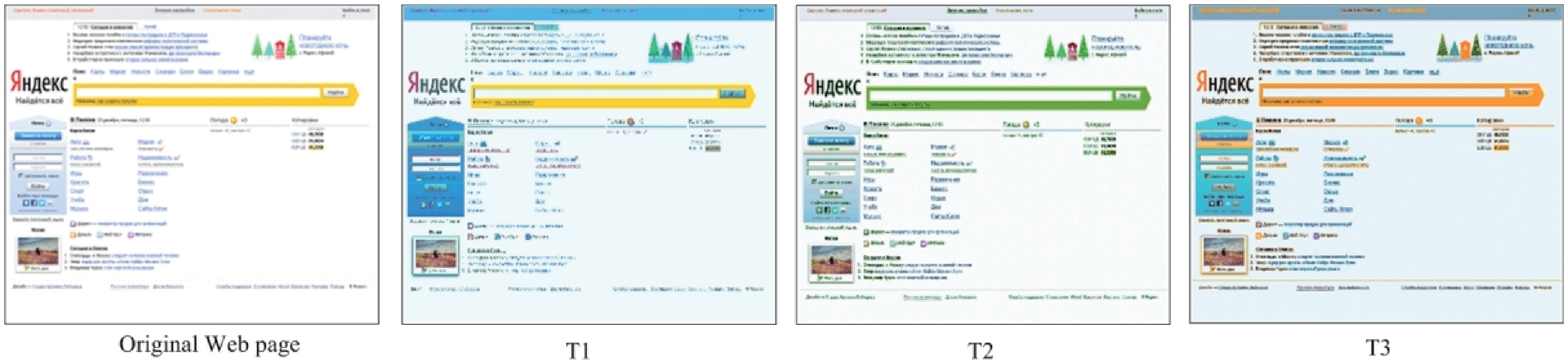}\\
 \vspace{-0.01in}\caption{The original Web page and three transferred Web pages. In our online user study, the average rating scores by 110 users of the two transferred pages (T1 and T2) are higher than that of the original page.}\vspace{-0cm}
\end{figure*}

\begin{figure*}[htbp]
\label{f3}
\centering
  \hspace{0in}\includegraphics[width= 17.76cm, height=1.386in]{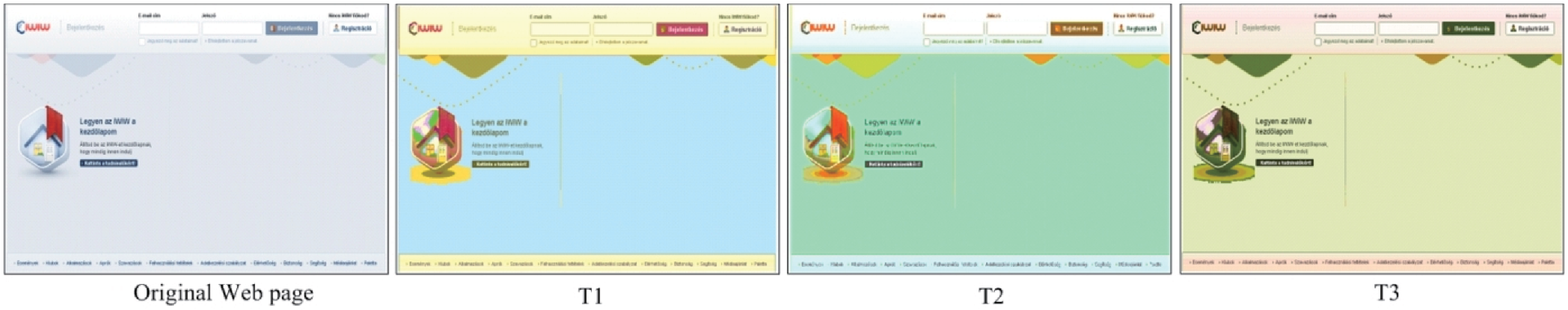}\\
 \vspace{-0.01in}\caption{The original Web page and three transferred Web pages. In our online user study, the average rating scores by 110 users of the two transferred pages (T1 and T2) are higher than that of the original page.}\vspace{0cm}
\end{figure*}

\begin{table}[t]
\begin{center}
\caption{Top 10 most correlated features}
\footnotesize
\begin{tabular}{|c | p{180pt} |c |}
\hline\hline
1 & The plan-fitting features of the lightness in CIELab & positive\\\hline
2&The mean of CIELab' s b dimension&negative\\\hline
3&The minimum difference between adjacent colors in terms of CHSV's V dimension&positive\\\hline
4&The second-largest difference between adjacent colors in terms of HSV's S dimension&negative\\\hline
5&The mean of RGB's B dimension&negative\\\hline
6&The sum-of-square error of the fitted 2D plane in CHSV&negative\\\hline
7&The largest difference between adjacent colors in terms of HSV's H dimension&negative\\\hline
8&The second-largest difference between adjacent colors in terms of CIELAB's L dimension&positive\\\hline
9&The minimum difference between adjacent colors in terms of CHSV's H dimension&negative\\\hline
10&The second-largest difference between adjacent colors in terms of RGB's B dimension&negative\\
\hline\hline
\end{tabular}
\label{tbl-1}
\end{center}
\end{table}
The correlation between features and color ratings of Web pages is also investigated on the 500 pages. The top 10 features that are most correlated to average color ratings are listed in Table 1. The lightness features are among the most important features for color assessment, which is consistent with the conclusions in \cite{Donovan:2011}. The blue color features are also important. The fourth feature is the brightness of the colors of a Web page. The minimum differences between adjacent colors are positively correlated to the color ratings. This is reasonable because a page with extreme lower difference between adjacent colors may result in poor readability. The 7th to 10th largest correlated features indicate that the differences between adjacent colors are important. The above observations indicate the differences between color ratings for Web pages and images.

\subsection{Case study for Web color transfer}
As described earlier, the reference Web pages in Web color transfer can be any forms of visual objects, including Web pages and images. We transfer colors for ten Web pages from the collected Web pages and some cartoon images. To assess the transferred Web pages, we launched a user study through the online user-study website in China \cite{Diaocha}. In our user study, each user was invited to rate the colors of ten groups of Web pages (including the pages in Figures 10-12). In each group, there are four pages that are made up of an original Web page and its three transferred result pages. Users rate each page in the range of ``1" to ``5". A total of 110 users (53 males and 57 females) rated the Web pages. About 76.4\% of the users surf the Internet more than 8 hours a week. About 94.54\% are in the age of 15 and 40.

According to the user study, 56.67\% of the transferred pages have higher rating scores compared with the original pages. Figures 10-12 show three color transfer examples. The leftmost page in each example is the original page, while the remaining three pages (i.e., T1, T2, and T3) are the transferred results whose scores are ordered top 3 by the learned color assessment model. Colors in some transferred pages have higher rating scores. For example, almost all the users in the age of 15 and 20 gave the highest scores to the T1 page in Fig. 12.

With our automatic Web color transfer and assessment framework, designers can obtain many Web-page images with different colors at very low costs. Although these images are not real Web pages, there are no obvious difference in color observation. These Web-page images demonstrate different color tones and perception to designers, which can provide new insights or inspiration to Web designers.

\subsection{Limitations}
The primary limitation is that the features are merely the color theme instead of the original colors of the fixed part. The secondary limitation is that this study assesses the colors of a Web page by simply assessing the compatibility of the color theme of that page. This simple strategy ignores the interactive nature of a Web page. For example, readability is a key property of Web pages \cite{Ling:2002}. However, it is ignored in this work, causing some poor-readability Web pages to be assessed with higher color scores. Another major limitation is the manually determination of the temporal parts of source Web pages. The manual operation may hinder the applications of our proposed Web color transfer and assessment approach in actual use.

\section{Conclusions}
Colors are very important to Web pages. This paper has investigated the assessment of Web page colors. A framework of assessment model construction has been proposed by learning existing online color theme-rating data for Web color assessment. Theories and techniques from Web mining, computer graphics, computer vision, and machine learning are introduced to address the main challenges in the key steps of the model construction (i.e., fixed part location, color theme extraction, and model learning). The constructed color assessment model is then applied into a new application, namely, color transfer for Web pages. Experiments and online user study results suggest the effectiveness of our proposed methodologies.


As a cross-discipline topic, our findings can promote the intersections between the involved disciplines. Color transfer between Web pages brings new changelings and may encourage the combination of more computer graphic techniques and Web mining techniques to Web design.


\bibliographystyle{abbrv}
\bibliography{template}
\appendix
\section{Salience map-based fixed part location}
This method reweights the proportion of colors according to the salience map output by salency detection\cite{Liu:2007}. The temporal parts usually have higher saliency weights, because temporal areas usually show image content or flash advertisement designed to attract users' attentions more than the fixed areas. On the contrary, the saliency of fixed parts is usually low. Based on these observations, a saliency-map based algorithm is proposed to locate the fixed part of a Web page. A pixel is considered as being in the fixed part if its salience value is below a certain threshold. This method is very sensitive to the results of saliency map detection. For many Web pages, areas of the fixed part are also detected as salency areas, and thus this method fails.
\section{Image synthesize-based fixed part location}
The colors of the temporal part are changed constantly, whereas those of the fixed parts remain relatively unchanged. Thus, for a specific Web page, we can first collect a number of temporal Web-page images at different times with the same URL of that page. Then the collected Web-page images are synthesized into a single image. As a consequence, we can extract color themes based on the synthesized image as the negative effects of colors from temporal parts have been reduced. The computational complexy of this method is high as the synthesized image has large size.
\balancecolumns
\end{document}